\documentclass[aps,prl,twocolumn,showpacs]{revtex4-1}
\usepackage{graphicx} 
\begin{document}
\title
{\bf Robust Ground State and Artificial Gauge in DQW Exciton Condensates under Weak Magnetic Field} 
\author{T. Hakio\u{g}lu$^{\bf (1,2)}$, Ege \"{O}zg\"{u}n$^{\bf (1)}$ and Mehmet G\"{u}nay$^{\bf (1)}$}
\affiliation{
${\bf (1)}$ {Department of Physics, Bilkent University, 06800 Ankara, Turkey}
and \break
${\bf (2)}$ {Institute of Theoretical and Applied Physics, 48740 Turun\c{c}, Mu\u{g}la, Turkey}
}
\begin{abstract}
Exciton condensate is a vast playground in studying a number of symmetries that are of high interest in the recent developments in topological condensed matter physics. In DQWs they pose highly nonconventional properties due to the pairing of non identical fermions with a spin dependent order parameter. Here, we demonstrate a new  feature in these systems: the robustness of the ground state to weak external B-field and the appearance of the artificial spinor gauge fields beyond a critical field strength where, negative energy pair-breaking quasi particle excitations are created in certain $k$ regions (DX-pockets). The DX-pockets are the Kramers symmetry broken analogs of the negative energy pockets examined in the 60s by Sarma, where they  principally differ from the latter in their non-degenerate energy bands due to the absence of the time reversal symmetry. They respect a disk or a shell-topology in $k$-space or a mixture between them depending on the B-field strength and the electron-hole mismatch. The Berry connection between the artificial flux and the TKNN number is made. The artificial spinor gauge field describes a collection of pure spin vortices in real space when the B-field has only inplane components.         

\end{abstract}
\maketitle
It has recently become clearer that fundamental symmetries play a much more subtle role in condensed matter physics. In particular, the interplay between the time reversal symmetry (TRS), spin rotation symmetry (SR), parity (P), particle-hole symmetry (PHS) lead into the theoretical and experimental discovery of an exotic zoo of topological insulators (TI)\cite{topo_ins}, topological superconductors (TSC)\cite{topo_SC} in one, two and three dimensions, and helped us in deeper understanding of the QHE and QSHE\cite{QHE,QSHE} within the periodic table of more general topological classes\cite{topo_class}. These structures once experimentally manipulated, are promising in devicing completely fault tolerant mechanisms for quantum computers\cite{topo_QComp}. In this field of research, the strong spin-orbit interaction with or without the magnetic field is the basic ingredient in providing the exotic topology in the momentum-spinor space\cite{SO_physics}.  

These fundamental symmetries that are important in TIs and TSCs, also play a subtle role in excitonic insulators not only in the normal phase of the exciton gas, but also in the  condensed phase in low temperatures. The basic difference from the PHS manifest TIs and the TSCs is that, the analogous symmetry in the excitonic systems, i.e. the fermion exchange (FX) symmetry is heavily broken. The absence of FX is minimally due to the different band masses and the orbital states of the electrons and holes and the parity breaking external E-field required in the experiments in order to prolong the exciton lifetime. Without the FX symmetry, the triplet and the singlet components have no definite parity and they can coexist within the same condensate. Additionally, despite the spin independence of the Coulomb interaction, exciton condensate (EC) breaks the spin degeneracy between the dark and the bright components from 4 to 2 due to the radiative exchange processes\cite{Combescott,TH_EO}. The four exciton spin states corresponding to the total spin-2 triplet (dark states) and the total spin-1 singlet (bright state) are connected by the TRS, imposing the condition on the spin dependent exciton order parameter: $\Delta_{\sigma \sigma^\prime}({\bf k})=-(-1)^{\sigma+\sigma^\prime}\Delta_{{\bar \sigma} {\bar \sigma}^\prime}^*(-{\bf k})$  where the dark and the bright states are the symmetric and antisymmetric combinations of the electron (hole) spins $\sigma (\sigma^\prime)=\{\pm 1/2\}$ respectively. Due to the real and isotropic Coulomb interaction, the order parameter matrix $\Delta_{\sigma \sigma^\prime}({\bf k})$ is real with vanishing offdiagonal triplet component, leaving two dark triplets $\Delta_{\sigma \sigma}({\bf k})$ and the bright singlet $\Delta_{\uparrow \downarrow}({\bf k})=-\Delta_{\downarrow \uparrow}({\bf k})$ nonzero. The breaking of FX implies that $\Delta_{\sigma \sigma^\prime}({\bf k})=-\Delta_{\sigma^\prime \sigma}(-{\bf k})$ is no longer respected\cite{sigrist-ueda}.  

The radiative exchange processes, inhibit the independent spin rotations of the electrons and holes in their own planes separating the dark and the bright contributions in magnitude. Considering these processes, we have recently confirmed that, the EC is dominated by the dark states\cite{TH_EO}. There are higher order weak mechanisms, known as Shiva diagrams\cite{Combescott} between two excitons, where the dark and the bright states can turn into each other by a fermion exchange. There are also intrinsic Dresselhaus as well as Rashba type spin-orbit couplings that are present already in many semiconductors. Nevertheless, the spin-orbit coupling in the case of EC is perturbatively smaller than the condensation energy gap\cite{TH_MAC} in comparison with the much stronger spin-orbit coupling in topologically interesting noncentrosymmetric superconductors. 
 
The manifestation/breaking of the TRS, SR, P and the FX symmetries play a fundamental role in the properties of the ground state of the EC. The physical parameters are, the exciton number $n_x$, the electron-hole number imbalance $n_-$ and the Coulomb interaction strength. The phase diagram is quite rich in that, the critical values of these parameters define a manifold even at zero temperature between the EC and the normal exciton gas\cite{TH_EO}. Within the condensed state, the Sarma I, II and the LOFF phases have been analytically examined by many authors in the context of atomic condensates\cite{atomic_condensates}. In exciton case the energy gap is inhomogeneous in ${\bf k}$ space due to long range Coulomb interaction and the numerical work is necessary to find under which conditions these different phases actually occur. We also report in this work that, both the Sarma-I and Sarma-II like phases\cite{Sarma} in ECs can be observed even when the Fermi surface mismatch is minimal, i.e. $n_-=0$. On the other hand, satisfying methods to search for the exotic LOFF phase require real space diagonalization and, up to our knowledge this has not been done yet for the ECs. Another high interest is the prediction of a strong critical Casimir force due to the strong dependence of the free energy on the layer separation near the phase boundary\cite{TH_EO}.

In this letter, we demonstrate another a new feature of the EC in response to a weak, adiabatically space dependent external B-field. That is the ground state topology and the appearance of artificial gauges in the real space created by these weak B-fields. In complimentary to the progress made in the {\bf k}-space TIs and TSCs, the search for artificial gauges has received significant attention in probing the real space topology of the neutral or charged atomic gases. In the particular case of neutral atoms, rotating a condensed atomic gas has been accomplished experimentally\cite{rotating_conds} by circularly polarized laser field and the apprearance of these gauge fields has been confirmed in the formation of superfluid vortices. Real space artificial pure gauge fields have been proposed based on the coupling of the internal quantum degrees of freedom with externally controllable adiabatic potentials\cite{geometric_pot}. 

Here we report that the real space adiabatic gauge fields can  be produced in the condensed excitonic background as a result of the absence of the FX symmetry. This symmetry is intrinsically broken due to the  electron-hole mass difference breaking the 4-fold spin degeneracy into a pair of Kramers doublets. The Kramers symmetry thus obtained is further broken with the application of the weak Zeeman field producing 4 non degenerate excitation bands. Two of these bands that are lowered by the Zeeman field can turn into the Sarma-I and II like bands beyond a critical magnetic field strength. A second method of strongly breaking the FX symmetry is by externally creating a number imbalance between the electrons and the holes. We examine in this article the consequences of both as well as their effects on the ground state topology.    
 
The electron-hole system in a typical semiconductor DQW structure is represented in the electron-hole basis $(\hat{e}_{{\bf k} \uparrow} \, \hat{e}_{{\bf k} \downarrow} \, \hat{h}^\dagger_{{\bf -k} \, \uparrow} \hat{h}^\dagger_{{\bf -k} \downarrow})$ by
\begin{eqnarray} 
{\cal H}=\pmatrix{{\tilde \epsilon}_{\bf k}^{(x)}\,\sigma_0 & \Delta^\dagger ({\bf k}) \cr 
                    \Delta ({\bf k}) & -{\tilde \epsilon}_{\bf k}^{(x)}\,\sigma_0\cr}+
{\tilde \epsilon}_{\bf k}^{(-)}\,\sigma_0 \times \sigma_0
\label{hamilt_1}
\end{eqnarray} 
where $\sigma_0$ is $2\times 2$ unit matrix, ${\tilde \epsilon}_{\bf k}^{(-)}=({\tilde \zeta}_{\bf k}^{(e)}-{\tilde \zeta}_{\bf k}^{(h)})/2$ is the mismatch energy and ${\tilde \epsilon}_{\bf k}^{(x)}=({\tilde \zeta}_{\bf k}^{(e)}+{\tilde \zeta}_{\bf k}^{(h)})/2$ with ${\tilde \zeta}_{\bf k}^{(e)}=\hbar^2k^2/(2m_e)-\mu_e, {\tilde \zeta}_{\bf k}^{(h)}=\hbar^2k^2/(2m_h)-\mu_h$ being the single particle energies (with the self energies) for the electrons and the holes with the masses $m_e$ and $m_h$, $\mu_e, \mu_h$ are their chemical potentials respectively and ${\bf \Delta}$ is a $2\times 2$ matrix representing the spin dependent order parameter\cite{sigrist-ueda}. 

This Hamiltonian can be diagonalized analytically, and the excitation spectra are $\lambda_{\bf k}=-{\tilde \epsilon}_{\bf k}^{(-)}+E_{\bf k}, \lambda_{\bf k}^*={\tilde \epsilon}_{\bf k}^{(-)}+E_{\bf k}$ where , $E_{\bf k}=\sqrt{({\tilde \epsilon}_{\bf k}^{(x)})^2+Tr[{\bf \Delta}({\bf k}){\bf \Delta}^\dagger({\bf k})]/2}$. Due to the time reversal symmetry, $\lambda_{\bf k}$ and $\lambda_{\bf k}^*$ are doubly degenerate. The excitations over the ground state can be described by the quasiparticle annihilation operators 
\begin{eqnarray}
\label{eigen_part1}
{\hat g}_{1,\bf k}&=&\alpha_{\bf k}\,\hat{e}_{{\bf k} \uparrow}+\beta_{\bf k}\,\hat{h}^\dagger_{{\bf -k} \, \uparrow}+\gamma_{\bf k}\,\hat{h}^\dagger_{{\bf -k} \, \downarrow} \nonumber \\ 
\\
{\hat g}_{2,\bf k}&=&\alpha_{\bf k}\,\hat{e}_{{\bf k} \downarrow}-\gamma_{\bf k}\,\hat{h}^\dagger_{{\bf -k} \, \uparrow}+\beta_{\bf k}\,\hat{h}^\dagger_{{\bf -k} \, \downarrow} \nonumber 
\end{eqnarray}
and 
\begin{eqnarray}
\label{eigen_part2}
{\hat g}_{3,\bf k}&=& \alpha_{\bf k}\,\hat{h}_{{\bf k} \uparrow}-\beta_{\bf k}\,\hat{e}^\dagger_{{\bf -k} \, \uparrow}+\gamma_{\bf k}\,\hat{e}^\dagger_{{\bf -k} \, \downarrow}    \nonumber \\
\\
{\hat g}_{4,\bf k}&=& \alpha_{\bf k}\,\hat{h}_{{\bf k} \downarrow}-\gamma_{\bf k}\,\hat{e}^\dagger_{{\bf -k} \, \uparrow}-\beta_{\bf k}\,\hat{e}^\dagger_{{\bf -k} \, \downarrow} \nonumber
\end{eqnarray} 
Here, $\alpha_{\bf k}={\cal C}_{\bf k}(E_{\bf k}+{\tilde \epsilon}_{\bf k}^{(x)})$, $\beta_{\bf k}={\cal C}_{\bf k}\Delta_{\uparrow \uparrow}({\bf k})$ and $\gamma_{\bf k}={\cal C}_{\bf k}\Delta_{\uparrow \downarrow}({\bf k})$ describe the normal, the dark and the bright condensate contributions in the ground state, where ${\cal C}_{\bf k}$ is determined by $\vert\alpha_{\bf k}\vert^2+\vert\beta_{\bf k}\vert^2+\vert\gamma_{\bf k}\vert^2=1$. 

\begin{figure}[b]
\includegraphics[scale=0.37,angle=-90]{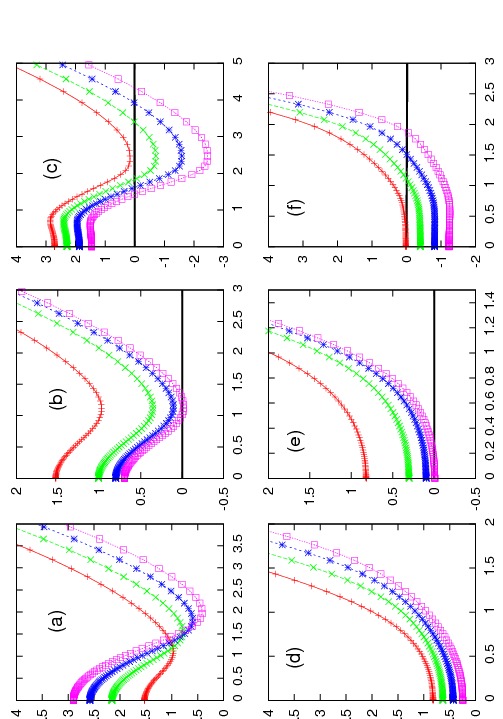}
\caption{(Color online) The upper ($\lambda_{\bf k}^{(-)}$) and the lower ($\lambda_{\bf k}^{*(-)}$) branches with negative Zeeman shifts are plotted as $k=\vert{\bf k}\vert$ varies (horizontal axes scaled by $a_B$) for various $n_x$ and $B$. The upper and lower branches: in (a,d) at $B=0$ with $n_x a_B^2=0.7, 0.5, 0.3, 0.1$ (from top to bottom at $k=0$); in (b,e) at $n_x a_B^2=0.1$ for $g^* B/B_0=0, 2.0, 2.8, 3.2$ (from top to bottom at $k=0$); in (c,f) at $n_x a_B^2=1.4$ for $g^* B/B_0=0,0.2,0.4,0.6$ (from top to bottom at $k=0$). The bands in (a,b,d,e) are doubly and in (c,f) singly degenerate. The zero of the vertical axes describes $E_G$.} \vspace{-0.5cm}
\label{k_crossings_2}
\end{figure}

\begin{figure}[t]
\includegraphics[scale=0.37,angle=-90]{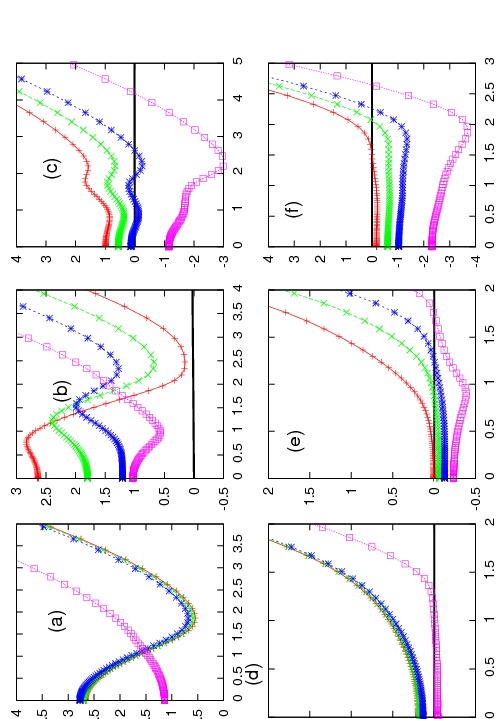}
\caption{(Color online) the same as Fig(\ref{k_crossings_2}) for $n_-\ne 0$. The upper and lower branches: (a,d) at $B=0$, $n_x a_B^2=0.55$ and $n_-=0,0.12,0.36,0.51$ (from top to bottom at $k=0$); (b,e) at $n_x=1.5$ and for $n_- a_B^2 =0,0.36,0.72,1.44$ at $B=0$ (from top to bottom at $k=0$); (c,f) at $n_x a_B^2=1.5$ and $n_-=1.1$ for $g^* B/B_0=0,0.2,0.4,1.0$ (from top to bottom at $k=0$).} \vspace{-0.5cm}
\label{k_crossings_3}
\end{figure}

In this article we ignore the effect of the radiative coupling and assume for simplicity that the dark and the bright pairing strengths are identical, i.e. $\vert\Delta_{\uparrow \uparrow}({\bf k})\vert=\vert\Delta_{\downarrow \downarrow}({\bf k})\vert=\vert \Delta_{\uparrow \downarrow}({\bf k})\vert$. Using the time reversal transformation for the real and isotropic order parameter  i.e. ${\hat \Theta}:\Delta_{\sigma \sigma}({\bf k})=\Delta_{{\bar \sigma} {\bar \sigma}}({\bf -k})=\Delta_{{\bar \sigma} {\bar \sigma}}({\bf k})$ and ${\hat \Theta}:\Delta_{\sigma {\bar \sigma}}({\bf k})=-\Delta_{{\bar \sigma} \sigma}({\bf -k})=-\Delta_{{\bar \sigma} \sigma}({\bf k})$ where $\sigma$ and ${\bar \sigma}$ are opposite spin orientations, it can be seen easily that 
\begin{eqnarray}
{\hat \Theta}: \Biggl[{{\hat g}_{{1 \choose 3},\bf k} \atop {\hat g}_{{2 \choose 4},\bf k}}\Biggr]=\Biggl[{~~{\hat g}_{{2 \choose 4},-\bf k} \atop -{\hat g}_{{1 \choose 3},\bf -k}}\Biggr]  
\label{TR_12}
\end{eqnarray}
Hence, Eq.(\ref{eigen_part1}) and (\ref{eigen_part2}) describe a pair of fermionic Kramers doublets. The ground state, described by $\vert \Psi_0 \rangle$ is annihilated by the operators in Eq.'s\,(\ref{eigen_part1}) and (\ref{eigen_part2}) and is given by $\vert \Psi_0 \rangle=\prod_{{\bf k}} \vert \psi_{{\bf k}}\rangle$ where $\vert \psi_{{\bf k}}\rangle=T_{\bf k}^{(1)}T_{\bf k}^{(2)}\vert 0\rangle$ are the {\it vacuum modes} with 
\begin{eqnarray}
\label{ground_state}
T_{\bf k}^{(1)}&=&\alpha_{\bf k}-\beta_{\bf k}\hat{e}^\dagger_{{\bf k} \,\uparrow}\hat{h}^\dagger_{{\bf -k} \,\uparrow}-\gamma_{\bf k} \hat{e}^\dagger_{{\bf k} \,\uparrow}\hat{h}^\dagger_{{\bf -k} \,\downarrow} \nonumber \\
\\
T_{\bf k}^{(2)}&=&\alpha_{\bf k}-\beta_{\bf k}\hat{e}^\dagger_{{\bf k} \,\downarrow}\hat{h}^\dagger_{{\bf -k} \,\downarrow}+\gamma_{\bf k} \hat{e}^\dagger_{{\bf k} \,\downarrow}\hat{h}^\dagger_{{\bf -k} \,\uparrow} \nonumber 
\end{eqnarray}
where ${\hat \Theta}:\vert \Psi_0 \rangle=\vert \Psi_0 \rangle$, hence the ground state is expectedly a time reversal singlet. The energy of the ground state is $E_G=-2\sum_{\bf k}\lambda_{\bf k}$ and the excitations are described by the Hamiltonian ${\cal H}^\prime=\sum_{\bf k}\lambda_{\bf k}^*[{\hat g}_{1,\bf k}^\dagger{\hat g}_{1,\bf k}+{\hat g}_{2,\bf k}^\dagger{\hat g}_{2,\bf k}]+\lambda_{\bf k}[{\hat g}_{3,\bf k}^\dagger{\hat g}_{3,\bf k}+{\hat g}_{4,\bf k}^\dagger{\hat g}_{4,\bf k}]$ where ${\cal H}^\prime={\cal H}-E_G$ is relative Hamiltonian with respect to the ground state.  We show the numerical self-consistent mean field solution of the energy bands in Fig(\ref{k_crossings_2} a,d) for $n_-=0$ and Fig(\ref{k_crossings_3} a,b,d,e) for finite $n_-$. Note that these bands are doubly degenerate where the corresponding eigenstates are related by time reversal. These are the non-conventional analogs of the disk shaped and the ring shaped bands that are studied first by Sarma in the 60s in the contect of conventional singlet superconductivity\cite{Sarma}.   

Once the condensate in Eq.\,(\ref{ground_state}) is formed with a negative condensation energy, a weak magnetic field is turned on as ${\bf B}({\bf r})=B_{\perp}{\hat e}_{\phi}+B_z {\hat e}_{z}$ where $B_z$ and $B_{\perp}$ are slowly spatially varying function of the radial coordinate $r=\vert{\bf r}\vert$ where ${\bf r}=(r,\phi)$. The field is weak firstly because we neglect the effect of the magnetic vector potential and that requires $\vert {\bf B}({\bf r})\vert \ll B_0$ where $B_0=\Phi_0 n_x$, with $\Phi_0$ as the flux quantum, is the critical field strength for Landau degeneracy. The second is that, we neglect the light hole influence on the heavy hole states\cite{SBMcD,Winkler}. 
The Zeeman coupling for the electron-heavy hole systems have been derived before\cite{SBMcD} as $U_z=-(\gamma_e {\bf \sigma^{(e)}}.{\bf B({\bf r})}+\gamma_h \sigma^{(h)}_z B_z)$ where $\gamma_i=g^* \mu_B^*/2$,  where $g^*$ is the effective g-factor\cite{g_factors} and $\mu_B^*=e\hbar/2m^*$ is the effective Bohr magneton with $m^*$ as the effective mass of the electron or the hole. Due to the intrinsic heavy-light hole splitting in the valence band (much larger than a typical Zeeman splitting), the Zeeman coupling for the heavy holes becomes highly anisotropic.  The Zeeman field breaks the Kramers symmetry between the quasiparticle operators in Eq.\,(\ref{eigen_part1}) and (\ref{eigen_part2}) as given in the block diagonal form of ${\cal H}^\prime$ as 
\begin{eqnarray}
Z=\pmatrix{A & B\cr B^* & -A\cr} 
\label{Zeeman_coupl_e-h}
\end{eqnarray}
where $A=\alpha_{\bf k}^2 B_z^{(e)}-(\beta_{\bf k}^2+\gamma_{\bf k}^2)B_z^{(h)}$, $B=\alpha_{\bf k}^2 B_{\perp}^{(e)}e^{-i\phi}$ in the $({\hat g}_{1,{\bf k}}, {\hat g}_{2,{\bf k}})$, whereas $A=-\alpha_{\bf k}^2 B_z^{(h)}+(\beta_{\bf k}^2+\gamma_{\bf k}^2)B_z^{(e)}$, $B=(\beta_{\bf k}^2+\gamma_{\bf k}^2)B_{\perp}^{(e)} e^{-i\phi}$ in the $({\hat g}_{3,{\bf k}}, {\hat g}_{4,{\bf k}})$ bases. Here we used $B_{z \atop \perp}^{({e \atop h})}=\gamma_{z \atop \perp}^{({e \atop h})} B_{z \atop \perp}$ for a compact notation. The excitation spectrum of ${\cal H}^\prime$ is split into $\lambda_{\bf k}^{(\pm)}=\lambda_{\bf k}\pm z_{\bf k}$ and $\lambda_{\bf k}^{*\,(\pm)}=\lambda_{\bf k}^*\pm z^*_{\bf k}$, where $z_{\bf k},z^*_{\bf k}=\sqrt{A^2+\vert B\vert^2}$. The Zeeman-shifted quasiparticles are 
\begin{eqnarray}
\Biggl[{{\hat G}_{{1 \choose 3},{\bf k}} \atop {\hat G}_{{2 \choose 4},{\bf k}}}\Biggr]&=&{\hat U}:\Biggl[{{\hat g}_{{1 \choose 3},{\bf k}} \atop {\hat g}_{{2 \choose 4},{\bf k}}}\Biggr] \nonumber \\
&=&\Biggl[{{\hat g}_{{1\choose 3},{\bf k}}\cos\frac{\theta_{\bf k}}{2}+{\hat g}_{{2\choose 4},{\bf k}}e^{-i\phi} \sin\frac{\theta_{\bf k}}{2} \atop  -{\hat g}_{{1\choose 3},{\bf k}}e^{i\phi} \sin\frac{\theta_{\bf k}}{2}+{\hat g}_{{2\choose 4},{\bf k}}\cos\frac{\theta_{\bf k}}{2}}\Biggr] 
\label{Zeeman_coupl_quasiparticles_1}
\end{eqnarray}
with $\tan\theta_{\bf k}=\vert B\vert/A$. The excitations in Eq.(\ref{Zeeman_coupl_quasiparticles_1}) are described by the Hamiltonian ${\cal H}^{\prime\prime}=\sum_{\bf k}[\lambda_{\bf k}^{*(+)}{\hat G}_{1,\bf k}^\dagger{\hat G}_{1,\bf k}+\lambda_{\bf k}^{*(-)}{\hat G}_{2,\bf k}^\dagger{\hat G}_{2,\bf k}+\lambda_{\bf k}^{(+)}{\hat G}_{3,\bf k}^\dagger{\hat G}_{3,\bf k}+\lambda_{\bf k}^{(-)}{\hat G}_{4,\bf k}^\dagger{\hat G}_{4,\bf k}]$. Unless, the {\it excitation energies} $\lambda_{\bf k}^{*(\pm)}, \lambda_{\bf k}^{(\pm)}$ are negative for some of the $k$-modes, the application of the Zeeman field does not change the ground state energy $E_G$ 
and the same ground state $\vert \Psi_0\rangle$ of the Hamiltonian ${\cal H}^\prime$ is now annihilated by the ${\hat G}_i$ operators. As the Zeeman energy is increased, the energy required to create an excitation in the first excited state becomes smaller and eventually at certain $k$ regions, $\lambda_{\bf k}^{*(-)}$ and(or) $\lambda_{\bf k}^{(-)}$ become(s) negative, creating a new ground state with energy lower than $E_G$. The numerical self consistent calculations for these branches with negative Zeeman shifts are shown in Fig.\,(\ref{k_crossings_2} b,c,e,f) for equal electron-hole concentrations, i.e. $n_-=0$, and in Fig.\,(\ref{k_crossings_3} c,f) for finite $n_-$. Since the Kramers symmetry is broken, we depicted only the relevant lower Zeeman branches in the figures.  

\begin{figure}
\includegraphics[scale=0.37,angle=-90]{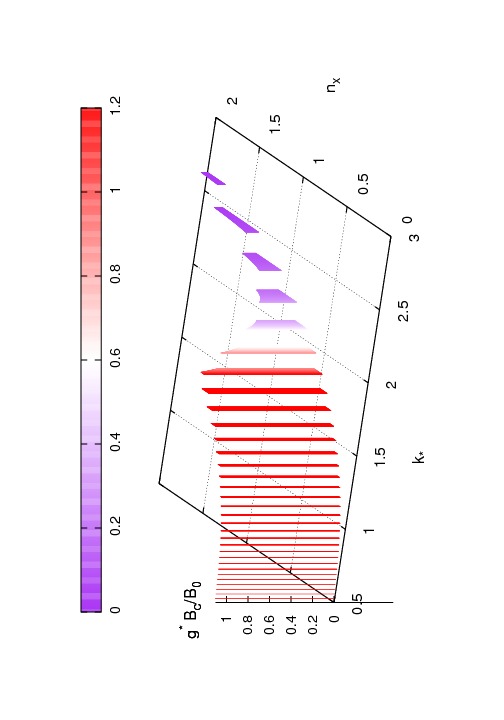}
\caption{(Color online) The critical B-field with their positions $k_* a_B$ as a function of the dimensionless exciton concentration $n_x a_B^2$. The colorbar measures the vertical scale.} \vspace{-0.5cm}
\label{k_crossing}
\end{figure}

At any arbitrary $\vert {\bf B}({\bf r})\vert\ll B_0$ exceeding the critical one, the condensate is represented by the new ground state 
\begin{eqnarray}
\vert\Psi_B\rangle=\prod_{\{k_*\}}{\hat G}_{{2,\bf k}_*}^\dagger \prod_{\{K_*\}} {\hat G}_{{4,\bf K}_*}^\dagger \vert\Psi_0\rangle
\label{new_GS}
\end{eqnarray}
where, $\{k_*\}$ and $\{K_*\}$ are the de-excitation pockets (DX-pockets) in the regions where $\lambda_{\bf k_*}^{*}<z_{\bf k_*}^*$ and $\lambda_{\bf K_*} < z_{\bf K_*}$ respectively. The DX-pockets correspond to one particle excitations with negative energy where breaking a pair by the ${\hat G}_{{2,\bf k}_*}^\dagger$ and ${\hat G}_{{4,\bf K}_*}^\dagger$ operations is energetically more favorable than keeping the pairs within the condensate. Those corresponding to $\lambda_{\bf k}^{(-)}$ branch have disk, i.e. $0<k<Q_1$, and those corresponding to the $\lambda_{\bf k}^{*(-)}$ branch have ring, i.e. $Q_2<k<Q_3$ topologies generating a rich spectrum of nonconventional ground states at different magnetic field strengths. The DX-pockets are shown in Fig.(\ref{k_crossings_2}.b,e) and (c,f) for the upper and the lower branch where they nearly touch $E_G$ in (b,e), and where they are given by the finite regions in (c,f) for different magnetic fields and concentrations. 
Before we discuss the appearance of the artificial gauge field, a justification is necessary for ignoring the magnetic vector potential. This is a good approximation when the magnetic field is considerably weaker than the critical field strength corresponding to the Landau level degeneracy at a fixed $n_x$ given as by $B_0=\Phi_0 n_x$ with $\Phi_0=h/e$ as the flux quantum. Considering the typical range $10^{10} \le n_x (cm^{-2})\le 10^{11}$, we have $0.4 \le B_0 (T) \le 4$. For the branch $\lambda_{\bf k_*}$ the critical field strength $B_c$ is found from $\lambda_{\bf k_*}=z_{\bf k_*}$ (i.e. $\lambda_{\bf k_*}^{(-)}=0$) which can be found as,  
\begin{eqnarray} 
g^*\frac{B_c}{B_0}=\frac{1}{\tilde{n_x}}\frac{(E_{{\bf k}_*}-{\tilde \epsilon}^{(x)}_{{\bf k}_*})/E_{Rd}}{\alpha_{{\bf k}_*}^2} 
\label{critical_ratio}
\end{eqnarray}  
where $\tilde{n_x}=\pi a_B^2 n_x$ is the dimensionless exciton concentration, $g^{*}=\sqrt{g_z^2+g_{\perp}^2}$ is the effective g-factor and $E_{Rd}=\hbar^2/(2m a_B^2)$ is the exciton Rydberg energy. The critical field on left hand side of Eq.\,(\ref{critical_ratio}) is defined by $B_c=\sqrt{(g_zB_z)^2+(g_{\perp}B_\perp)^2}/g^{*}$ and we assumed for simplicity that $B_z=B_\perp$. We can roughly estimate $B_c/B_0$ using $g^*\simeq -3$, $m^*\simeq 0.067 m_e$, where $m_e$ is the electron mass in vacuum, for $n_x a_B^2=1$ by ignoring the self energy corrections to ${\tilde \epsilon}^{(x)}_{{\bf k}_*}$. as the B-field is increased, the earliest de-excitation occurs at the point where gap is the weakest, $E_{k_*}\simeq \mu_x$, where $\mu_x\simeq n_x/2\Gamma$, with $\Gamma$ being the two dimensional density of states, we find that
\begin{eqnarray} 
B_c \simeq \frac{2n_x}{g^* \Gamma \mu_B^*}
\label{estimate_B_c}
\end{eqnarray}  
where the coefficient $2/g^* \Gamma \mu_B^*$ is in flux units, and a simple calculation yields that $2/g^* \Gamma \mu_B^*=(4/g^*)\Phi_0$ with $B_c=4/g^* \, n_x\Phi_0$. This result, which is a comparison between the Zeeman energy and the Landau level splitting is quite expected and verifies that Eq.\,(\ref{critical_ratio}) yields the expected result in the weak condensate limit. The numerical result for $g^*\frac{B_c}{B_0}$ in Eq.(\ref{critical_ratio}) for the disk shaped DX-pockets is plotted for various concentrations in Fig.(\ref{k_crossing}). We believe that In-based semiconductors with a large $g^*$ factor are good candidates to observe the DX-pockets. 

An important result here is the emergence of an artificial gauge field for $B_c \le B$ as given by ${\bf A}({\bf r})=-i\hbar\langle\Psi_B\vert  \nabla_{\bf r} \vert\Psi_B\rangle$. Using Eq.(\ref{new_GS}) we find 
\begin{eqnarray}
{\bf A}({\bf r})=-\frac{\hbar {\hat e}_{\phi}}{r} \sum_{{k \in \atop \{k_*\}+\{K_*\}}} \sin^2\frac{\theta_{\bf k}}{2}
\label{artificial_gauge_field_gs} 
\end{eqnarray}
which is an overall pure gauge field present only for those modes in the DX-pockets. In deriving Eq.(\ref{artificial_gauge_field_gs}) we ignored the $\vert{\bf r}\vert$ dependence of $\theta_{\bf k}$ through $B_\perp$ which is experimentally justified considering the microscopic size of the condensate. Due to the dependence of $\theta_{\bf k}$ on the ratio $B_\perp/B_z$, the magnetic field dependence of ${\bf A}({\bf r})$ mainly comes from the boundaries of the DX-pockets. 

Since the boundaries of the DX-pockets (denoted by $\Sigma_{\bf K_*}, \Sigma_{\bf k_*}$) are defined by where the excitation gap closes, i.e. $\lambda_{\bf k_*}^*-z_{\bf k_*}^*=0$ and $\lambda_{\bf K_*}-z_{\bf K_*}=0$, it is appealing to know if a non-trivial topology is present in the band structure and whether there is any connection with the artificial vortex in Eq.(\ref{artificial_gauge_field_gs}). Due to the slowly varying magnetic field, and for a given ground state mode {\bf k}, it is suggested by Eq.(\ref{Zeeman_coupl_e-h}) that, this topology is present not in the spinor-{\bf k}, but in the spinor-{\bf r} space. 
Generalizing the topological index by TKNN\cite{TKNN} in the form,   
\begin{eqnarray}
I=\sum_{{k \in \atop \{k_*\}+\{K_*\}}} \int d{\bf \ell}_{\bf r} . \sum_{\lambda} \langle \chi_\lambda({\bf k})\vert \nabla_{\bf r} \vert \chi_\lambda({\bf k})\rangle 
\label{TKNN_number}
\end{eqnarray}
where $d\ell_{\bf r}$ describes the real-space line integral, $\vert \chi_\lambda \rangle$ is the eigenstate of Eq.(\ref{Zeeman_coupl_e-h}) in the two spin eigen configurations $\lambda$ corresponding to the Zeeman lowered energy band yielding the DX-pocket, it can be seen that $I_{TKNN}$ is nothing but the total artificial flux enclosed within the DX-regions in Eq.(\ref{artificial_gauge_field_gs}). If $B_z=0$, then $\theta_{\bf k}=\pi/2$ and Eq.(\ref{artificial_gauge_field_gs}) describes a spin vortex, i.e. ${\bf A}({\bf r})=-(N\hbar/2r)\,{\hat e}_{\phi}$. In this case, every single mode in the disk or ring shaped DX-pockets carries $\hbar/2$ flux quantum with an integer $I$ number equal to the total number of modes in the DX-pockets $\{k_*\}+\{K_*\}$.     

Exotic properties are being studied extensively in the topology of the energy bands of the insulators, superconductors as well as their interfaces where the external magnetic field and the spin-orbit coupling play essential role with correlated spin and momentum configurations. These systems are composed of a single particle species with or without spin degrees of freedom with manifest particle-hole symmetry but a broken time reversal in the former whereas manifest in the latter. The FX symmetry is the analog of the particle-hole symmetry and, in ECs, with two species of paired particles, it is broken, hence no doubling issues arise for the fermion degree of freedom. In the model studied here, contrary to the particle-hole symmetric superconductors with violated parity, the appearance of the triplet and the singlet condensates with mixed parities is the result of the FX symmetry breaking which leads to a real space topology in the presence of a textured B-field. Spinor related Fermi space topology has been recently detected in the spin-ARPES measurements\cite{spin_ARPES}. We believe that this technique with an additional Fourier decomposition can also be applied to the real space-spinor topology studied here.       


\begin{thebibliography}{99}
\bibitem{g_factors} In quantum wells, the effective g-factors depend on the well width. Typical values can be found in (\cite{Winkler}) as $g_z\simeq -0.2$ and $g_{\perp}=-0.05$ for a $70 \AA$ wide well and for $GaAs-Al_{x}Ga_{1-x}As$.
\end{thebibliography}
\end{document}